\documentstyle[prd,aps,preprint,tighten,epsfig]{revtex}

\begin{document}

\draft

\title{Nontrivial Correlation between the CKM and MNS Matrices}
\author{\bf Zhi-zhong Xing}
\address{CCAST (World Laboratory), P.O. Box 8730, Beijing 100080, China \\
and Institute of High Energy Physics, Chinese Academy of Sciences, \\
P.O. Box 918 (4), Beijing 100039, China
\footnote{Mailing address} \\
({\it Electronic address: xingzz@mail.ihep.ac.cn}) }
\maketitle

\begin{abstract}
We point out that the Cabibbo-Kobayashi-Maskawa (CKM) quark mixing
matrix $V_{\rm CKM}$ and the Maki-Nakagawa-Sakata (MNS) lepton
mixing matrix $V_{\rm MNS}$ can naturally be correlated in a class
of seesaw models with grand unification, but the texture of their
correlation matrix ${\cal F}_\nu$ is rather nontrivial. The
bimaximal mixing pattern of ${\cal F}_\nu$ is disfavored by
current data, and other special forms of ${\cal F}_\nu$ may suffer
from fine-tuning of the free phase parameters in fitting the
so-called quark-lepton complementarity relation. A straightforward
calculation of ${\cal F}_\nu$ in terms of $V_{\rm CKM}$ and
$V_{\rm MNS}$ reveals a striking feature of ${\cal F}_\nu$: its
(1,3) element cannot be zero or too small, no matter whether the
(1,3) elements of $V_{\rm CKM}$ and $V_{\rm MNS}$ are vanishing or
not. We also add some brief comments on possible radiative
corrections to $V_{\rm CKM}$ and $V_{\rm MNS}$.
\end{abstract}


\newpage

\framebox{\large\bf 1} ~ Recent solar \cite{SNO}, atmospheric
\cite{SK}, reactor (KamLAND \cite{KM} and CHOOZ \cite{CHOOZ}) and
accelerator (K2K \cite{K2K}) neutrino oscillation experiments have
provided us with very robust evidence that neutrinos are massive
and lepton flavors are mixed. The phenomenon of lepton flavor
mixing can be described by a $3\times 3$ unitary matrix $V_{\rm
MNS}$, commonly referred to as the Maki-Nakagawa-Sakata (MNS)
matrix \cite{MNS}. Current experimental data indicate that $V_{\rm
MNS}$ involves two large angles ($\theta_{12} \sim 33^\circ$ and
$\theta_{23} \sim 45^\circ$) and one small angle ($\theta_{13}
\lesssim 9^\circ$) in the standard parametrization. The magnitude
of $\theta_{13}$ remains unknown, but a global analysis of the
presently available neutrino oscillation data \cite{FIT} hints
that $\theta_{13} \sim 3^\circ$ seems to be most likely. On the
other hand, three nontrivial CP-violating phases of $V_{\rm MNS}$
(denoted by $\delta_{\rm MNS}$, $\rho$ and $\sigma$) are entirely
unrestricted. One particularly important target of the future
neutrino experiments is just to measure $\theta_{13}$ and
$\delta_{\rm MNS}$ (of Dirac type) and to constrain $\rho$ and
$\sigma$ (of Majorana type).

In comparison with the MNS matrix $V_{\rm MNS}$, the
Cabibbo-Kobayashi-Maskawa (CKM) quark flavor mixing matrix $V_{\rm
CKM}$ involves three small mixing angles and one large
CP-violating phase ($\vartheta_{12} \approx 13^\circ$,
$\vartheta_{23} \approx 2.4^\circ$, $\vartheta_{13} \approx
0.2^\circ$ and $\delta_{\rm CKM} \sim 65^\circ$ in the standard
parametrization \cite{PDG}). The apparent difference between the
CKM and MNS matrices requires a good dynamical reason in a
fundamental theory of flavor mixing and CP violation, in
particular when an underlying lepton-quark symmetry is concerned.
Some phenomenological speculation about possible relations between
$\theta_{ij}$ and $\vartheta_{ij}$, such as
\cite{Raidal,Smirnov,Kang}
\begin{equation}
\theta_{12} + \vartheta_{12} \; \approx \; 45^\circ \;
\end{equation}
and \cite{Raidal}
\footnote{Note that $\theta_{23} - \vartheta_{23} \approx
45^\circ$ can indistinguishably be expected from current
experimental data.}
\begin{equation}
\theta_{23} + \vartheta_{23} \; \approx \; 45^\circ \; ,
\end{equation}
has appeared
\footnote{Note that $\theta_{13} \sim \vartheta_{13}$ was also
mentioned in Ref. \cite{Raidal}. However, we find that the present
best-fit value of $\theta_{13}$ favors $\theta_{13} \sim
\vartheta_{23}$ instead of $\theta_{13} \sim \vartheta_{13}$.}.
If such relations could survive the test with more accurate
experimental data in the near future, would they be just
accidental or imply a kind of lepton-quark symmetry? The latter is
certainly attractive to theorists, although it remains unclear
what symmetry exists between leptons and quarks.

The main purpose of this paper is to investigate a simple but
natural relation between $V_{\rm MNS}$ and $V_{\rm CKM}$ for a
large class of seesaw models with grand unification,
\begin{equation}
V_{\rm MNS} \; =\; V_{\rm CKM}^\dagger {\cal F}_\nu \; ,
\end{equation}
in which ${\cal F}_\nu$ denotes a unitary matrix associated with
the diagonalization of the effective neutrino mass matrix $M_\nu$
\cite{Ramond}. We shall show that it is phenomenologically
disfavored for ${\cal F}_\nu$ to take the bimaximal mixing form,
even if the nontrivial phase effects in ${\cal F}_\nu$ are taken
into account. This observation is contrary to some previous
arguments that the CKM matrix might just measure the deviation of
the MNS matrix from exact bimaximal mixing
\cite{Raidal,Smirnov,Giunti}. Furthermore, we point out that a
slight modification of the bimaximal mixing pattern of ${\cal
F}_\nu$ will allow us to reproduce the so-called quark-lepton
complementarity relation in Eq. (1), but the fine-tuning of
relevant unknown parameters seems unavoidable. A straightforward
calculation of ${\cal F}_\nu$ in terms of the mixing angles and
CP-violating phases of $V_{\rm CKM}$ and $V_{\rm MNS}$ reveals a
striking feature of ${\cal F}_\nu$: its (1,3) element cannot be
vanishing or too small, no matter whether $\theta_{13}$ and
$\vartheta_{13}$ are taken to be zero or not. Therefore we
conclude that the texture of ${\cal F}_\nu$ is rather nontrivial
in the seesaw models. We also make some brief comments on possible
quantum corrections to fermion mass matrices and their threshold
effects on the flavor mixing parameters that are related to one
another by Eq. (3).

\vspace{0.3cm}

\framebox{\large\bf 2} ~
First of all, let us explain why Eq. (3) naturally holds for a class of
seesaw models with lepton-quark unification. It is well-known that the
CKM matrix $V_{\rm CKM} \equiv V^\dagger_{\rm u} V_{\rm d}$ arises from
the mismatch between the diagonalization of the up-type quark mass matrix
$M_{\rm u}$ and that of the down-type quark mass matrix $M_{\rm d}$:
\begin{eqnarray}
M_{\rm u} & ~ = ~ & V_{\rm u} ~ \overline{M}_{\rm u} ~ U^\dagger_{\rm u} \; ,
\nonumber \\
M_{\rm d} & = & V_{\rm d} ~ \overline{M}_{\rm d} ~ U^\dagger_{\rm d} \; ,
\end{eqnarray}
where $\overline{M}_{\rm u} = {\rm Diag}\{m_u, m_c, m_t\}$,
$\overline{M}_{\rm d} = {\rm Diag}\{m_d, m_s, m_b\}$,
$U_{\rm u,d}$ and $V_{\rm u,d}$ are unitary matrices. Similarly,
the MNS matrix $V_{\rm MNS} \equiv V^\dagger_l V_\nu$ comes from the
mismatch between the diagonalization of the charged lepton mass matrix
$M_l$ and that of the (effective) Majorana neutrino mass matric $M_\nu$:
\begin{eqnarray}
M_l & ~ = ~ & V_l ~ \overline{M}_l ~ U^\dagger_l \; ,
\nonumber \\
M_\nu & = & V_\nu ~ \overline{M}_\nu ~ V^T_\nu \; ,
\end{eqnarray}
in which $\overline{M}_l = {\rm Diag}\{m_e, m_\mu, m_\tau\}$,
$\overline{M}_\nu = {\rm Diag}\{m^{~}_1, m^{~}_2, m^{~}_3\}$,
$V_{l,\nu}$ and $U_l$ are unitary matrices. Taking account of the
canonical (Type I) seesaw mechanism \cite{Minkowski}, we have
\begin{equation}
M_\nu = M_{\rm D} M^{-1}_{\rm R} M^T_{\rm D} \; , ~
\end{equation}
where $M_{\rm D}$ and $M_{\rm R}$ stand respectively for the Dirac
neutrino mass matrix and the heavy (right-handed) Majorana
neutrino mass matrix. A diagonalization of $M_{\rm D}$ is
straightforward:
\begin{equation}
M_{\rm D} =  V_{\rm D} ~ \overline{M}_{\rm D} ~ U^\dagger_{\rm D} \; , ~
\end{equation}
where $\overline{M}_{\rm D} = {\rm Diag}\{m_x, m_y, m_z\}$ with
$m_x$, $m_y$ and $m_z$ being the (positive) eigenvalues of $M_{\rm
D}$, $V_{\rm D}$ and $U_{\rm D}$ are unitary matrices. Since the
idea of SO(10) grand unification \cite{SO10} provides a natural
framework in which leptons and quarks are correlated and the
seesaw mechanism automatically works
\footnote{It is worth pointing out that the non-canonical (Type
II) seesaw mechanism works more naturally in most of the realistic
SO(10) models \cite{Bajc}. For simplicity, here we assume that the
canonical seesaw relation in Eq. (6) holds as a leading-order
approximation of the non-canonical seesaw relation.},
we just concentrate on it in the following. But we restrict
ourselves to a phenomenologically simplified and interesting
scenario: $M_{\rm D} = M_{\rm u}$ and $M_l = M_{\rm d}$, and all
of them are symmetric \cite{BW}. We are then left with $V_{\rm D}
= U^*_{\rm D} = V_{\rm u} = U^*_{\rm u}$ and $V_l = U^*_l = V_{\rm
d} = U^*_{\rm d}$, as well as $m_x = m_u$, $m_y = m_c$ and $m_z =
m_t$. A relation between the MNS matrix $V_{\rm MNS}$ and the CKM
matrix $V_{\rm CKM}$ turns out to be
\begin{equation}
V_{\rm MNS} \; =\; V^\dagger_{\rm d} V_\nu \; =\;
V^\dagger_{\rm CKM} \left (V^\dagger_{\rm u} V_\nu \right ) \; =\;
V^\dagger_{\rm CKM} {\cal F}_\nu \; ,
\end{equation}
where ${\cal F}_\nu \equiv V^\dagger_{\rm u} V_\nu$. With the help
of the seesaw relation in Eq. (6), the role of ${\cal F}_\nu$ can
be seen clearly:
\begin{equation}
{\cal F}_\nu \overline{M}_\nu {\cal F}^T_\nu \; = \;
\overline{M}_{\rm u} V^T_{\rm u} M^{-1}_{\rm R} V_{\rm u}
\overline{M}_{\rm u} \; .
\end{equation}
In other words, the unitary matrix ${\cal F}_\nu$ transforms the
combination $\overline{M}_{\rm u} V^T_{\rm u} M^{-1}_{\rm R}
V_{\rm u} \overline{M}_{\rm u}$ into the diagonal (physical) mass
matrix of three light neutrinos. Given ${\cal F}_\nu$ and $V_{\rm
u}$, the heavy Majorana neutrino mass matrix $M_{\rm R}$ is
determined by
\begin{equation}
M_{\rm R} \; =\; V_{\rm u} \overline{M}_{\rm u} {\cal F}^*_\nu
\overline{M}_\nu^{-1} {\cal F}^\dagger_\nu \overline{M}_{\rm u}
V^T_{\rm u} \; .
\end{equation}
This {\it inverted} seesaw relation indicates that the mass scale
of three right-handed neutrinos is roughly $m^2_t/m^{~}_3$, if the
light neutrino mass spectrum is essentially hierarchical.

We argue that Eq. (8) is a natural relation between CKM and MNS
matrices in a class of seesaw models with lepton-quark unification.
It will not be favored by current neutrino oscillation data, however,
if ${\cal F}_\nu$ takes the bimaximal mixing pattern.

\vspace{0.3cm}

\framebox{\large\bf 3} ~ Because ${\cal F}_\nu$ depends on both
$M_{\rm R}$ and $V_{\rm u}$ through Eq. (9), it is in general a
complex unitary matrix. The most generic form of ${\cal F}_\nu$
with bimaximal mixing can be written as
\begin{equation}
{\cal F}_\nu \;=\; P_\nu \left ( \matrix{
\frac{1}{\sqrt{2}} & \frac{1}{\sqrt{2}} & 0 \cr
- \frac{1}{2} & \frac{1}{2} & \frac{1}{\sqrt{2}} \cr
\frac{1}{2} & - \frac{1}{2} & \frac{1}{\sqrt{2}} \cr} \right )
Q_\nu \; ,
\end{equation}
where $P_\nu \equiv {\rm Diag}\{e^{i\phi^{~}_x}, e^{i\phi^{~}_y},
e^{i\phi^{~}_z}\}$ and $Q_\nu \equiv {\rm Diag}\{e^{i\phi^{~}_1},
e^{i\phi^{~}_2}, e^{i\phi^{~}_3}\}$ are two phase matrices. Taking
account of Eq. (8), one may choose two independent phases (or
their combinations) of $Q_\nu$ as the Majorana-type CP-violating
phases of $V_{\rm MNS}$. On the other hand, the phases of $P_\nu$
can affect both the mixing angles and the Dirac-type CP-violating
phase of $V_{\rm MNS}$. To see this point in a more transparent way,
we make use of the standard parametrizations of $V_{\rm CKM}$ and
$V_{\rm MNS}$ \cite{Xing04},
\begin{eqnarray}
V_{\rm CKM} & = & R_{23}(\vartheta_{23}) \otimes
\Gamma_\delta(\delta_{\rm CKM}) \otimes R_{13}(\vartheta_{13})
\otimes R_{12}(\vartheta_{12}) \; ,
\nonumber \\
V_{\rm MNS} & = & R_{23}(\theta_{23}) \otimes
\Gamma_\delta(\delta_{\rm MNS}) \otimes R_{13}(\theta_{13})
\otimes R_{12}(\theta_{12}) \otimes Q_\nu \; ,
\end{eqnarray}
in which $R_{ij}$ denotes the rotation matrix in the ($i,j$)-plane
with the mixing angle $\theta_{ij}$ or $\vartheta_{ij}$ (for $ij =
12, 23, 13$), and $\Gamma_\delta(\delta) \equiv {\rm Diag}\{1, 1,
e^{i\delta}\}$ is a phase matrix consisting of the Dirac phase of
CP violation. When Eq. (12) is applied to Eq. (8), a proper
rephasing of three charged lepton fields has been implied for the
sake of phase consistency on the two sides of Eq. (8)
\footnote{A general parametrization of $V_{\rm CKM}$ should
include two (unobservable) phase matrices on its left-hand side
(denoted as $P$) and its right-hand side (denoted as $Q$). Taking
account of Eq. (8), we find that $P^\dagger$ can be absorbed into
$P_\nu$ of ${\cal F}_\nu$, while $Q^\dagger$ can be removed by a
redefinition of the non-physical phases of three charged lepton
fields. Therefore, the standard parametrizations of $V_{\rm CKM}$
and $V_{\rm MNS}$ given in Eq. (12) together with the most generic
form of ${\cal F}_\nu$ taken in Eq. (11) are consistent with Eq.
(8) and the Lagrangian of lepton and quark Yukawa interactions.}.
We are then allowed to calculate three mixing angles of $V_{\rm
MNS}$ in terms of those of $V_{\rm CKM}$. The approximate results
are
\footnote{To leading order in our approximation, another
(unobservable) phase combination $(\phi_z -\phi_y)$ does not
appear in Eqs. (13) and (14).}
\begin{eqnarray}
\tan\theta_{12} & \approx & \frac{\sqrt{1 - \sqrt{2}\tan\vartheta_{12}
\cos (\phi_y - \phi_x)}}
{\sqrt{1 + \sqrt{2}\tan\vartheta_{12} \cos (\phi_y - \phi_x)}} \; ,
\nonumber \\
\tan\theta_{23} & \approx & \cos\vartheta_{12} \; ,
\nonumber \\
\sin\theta_{13} & \approx & \frac{1}{\sqrt{2}}\sin\vartheta_{12} \; ,
\end{eqnarray}
where the strong hierarchy of three quark mixing angles has been
taken into account. In addition, the Jarlskog invariant of
CP violation \cite {J} defined for $V_{\rm MNS}$ is given by
\begin{equation}
{\cal J}_{\rm MNS} \; \approx \; \frac{1}{8\sqrt{2}} \sin 2\vartheta_{12}
\sin (\phi_y - \phi_x) \;
\end{equation}
to the leading order, only if the phase difference $(\phi_y
-\phi_x)$ is not too close to $0$ or $\pm\pi$. Then the
maximally-allowed magnitude of ${\cal J}_{\rm MNS}$ is $|{\cal
J}_{\rm MNS}| \approx 4\%$ for $(\phi_y -\phi_x) \approx \pm
\pi/2$. But this special phase condition is not favored by Eq.
(13), because it will lead to $\tan\theta_{12} \approx 1$ or
$\theta_{12} \approx 45^\circ$. Taking account of ${\cal J}_{\rm
MNS} = \sin\theta_{12}\cos\theta_{12}\sin\theta_{23}
\cos\theta_{23}\sin\theta_{13}\cos^2\theta_{13}\sin\delta_{\rm
MNS}$, we are then able to determine the Dirac phase $\delta_{\rm
MNS}$ from Eqs. (13) and (14). We obtain $\delta_{\rm MNS} \approx
(\phi_y - \phi_x)$ in the leading-order approximation. Therefore
the phase matrix $P_\nu$ of ${\cal F}_\nu$, which may
significantly affect $\theta_{12}$ and $\delta_{\rm MNS}$ of
$V_{\rm MNS}$, should not be ignored.

Two more comments on the consequences of Eqs. (13) and (14) are in order.

(a) At first glance, it is absolutely impossible to reproduce the
empirical relations given in Eqs. (1) and (2) from Eq. (13),
simply because of the existence of unknown $(\phi_x, \phi_y,
\phi_z)$ phases. It is worth emphasizing that the special
assumption $\phi_x = \phi_y = \phi_z =0$ taken in the literature
is not justified, since ${\cal F}_\nu$ is naturally expected to be
complex in the seesaw models. For this reason, we argue that Eqs.
(1) and (2) are most likely to be a numerical accident.

(b) Even if $\phi_x = \phi_y = \phi_z =0$ is assumed, it will be
difficult to straightforwardly derive Eq. (1) from Eq. (13). In
this case, the expression of $\tan\theta_{12}$ in Eq. (13) is
simplified to
\begin{equation}
\tan\theta_{12} \; \approx \; \frac{\sqrt{1 - \sqrt{2}\tan\vartheta_{12}}}
{\sqrt{1 + \sqrt{2}\tan\vartheta_{12}}} \; ,
\end{equation}
Typically taking $\vartheta_{12} \approx 13^\circ$, we obtain
$\theta_{12} \approx 35^\circ$ from Eq. (15). In addition,
$\theta_{23} \approx 44^\circ$ and $\theta_{13} \approx 9^\circ$
can be obtained from Eq. (13). It turns out that the sum of
$\theta_{12}$ and $\vartheta_{12}$ amounts to $48^\circ$, which is
not perfectly in agreement with Eq. (1). On the other hand, the
prediction $\theta_{13} \approx 9^\circ$ is too close to the
present experimental upper limit and far away from the best-fit
result (namely, $\theta_{13} \sim 3^\circ$ \cite{FIT}). Note also
that the assumption $\phi_x = \phi_y = \phi_z =0$ leads to tiny CP
violation in the lepton sector:
\begin{equation}
|{\cal J}_{\rm MNS}| \; \approx \;
\frac{1}{4\sqrt{2}} \sin\vartheta_{13}\sin\delta_{\rm CKM} \; ,
\end{equation}
where only the leading term has been given. Numerically, $|{\cal
J}_{\rm MNS}| \sim 5 \times 10^{-4}$, too small to be measured in
the future long-baseline neutrino oscillation experiments
\cite{Zhang}.

In order to successfully achieve $\theta_{12} + \vartheta_{12}
\approx 45^\circ$ from $V_{\rm MNS} = V^\dagger_{\rm CKM} {\cal
F}_\nu$, a straightforward (and somehow trivial) way is to modify
the bimaximal mixing pattern of ${\cal F}_\nu$. For illustration,
let us consider the following form of ${\cal F}_\nu$:
\begin{equation}
{\cal F}_\nu = P_\nu \left ( \matrix{ c & s & 0 \cr -
\frac{1}{\sqrt 2}s & \frac{1}{\sqrt 2}c & \frac{1}{\sqrt{2}} \cr
\frac{1}{\sqrt 2}s & - \frac{1}{\sqrt 2}c & \frac{1}{\sqrt{2}}
\cr} \right ) Q_\nu \; ,
\end{equation}
where $c\equiv \cos\theta$ and $s\equiv \sin\theta$ with $\theta$
being an unspecified angle. Eq. (11) can obviously be reproduced
from Eq. (17) by taking $\theta = 45^\circ$. Taking account of the
correlation between $V_{\rm MNS}$ and $V_{\rm CKM}$ in Eq. (3) or
(8), we now arrive at
\begin{eqnarray}
\tan\theta_{12} & \approx & \frac{\sqrt{\tan\theta -
\sqrt{2}\tan\vartheta_{12} \cos (\phi_y - \phi_x)}}
{\sqrt{\cot\theta + \sqrt{2}\tan\vartheta_{12} \cos (\phi_y -
\phi_x)}} \;
\nonumber \\
{\cal J}_{\rm MNS} & \approx & \frac{1}{8\sqrt{2}} \sin 2\theta
\sin 2\vartheta_{12} \sin (\phi_y - \phi_x) \; ,
\end{eqnarray}
in the leading-order approximation. The results for
$\tan\theta_{23}$ and $\sin\theta_{13}$ obtained in Eq. (13) keep
unchanged. The relation $\theta_{12} + \vartheta_{12} \approx
45^\circ$, if it holds, implies a kind of correlation between
$\theta$ and $(\phi_y - \phi_x)$ as restricted by Eq. (18). To be
more concrete, we find that this correlation reads as
\begin{equation}
\cos (\phi_y - \phi_x) \; \approx \; \frac{2 \tan\vartheta_{12} -
\cos 2\theta}{\sqrt{2} \tan\vartheta_{12} \sin 2\theta} \;
\end{equation}
to the lowest order. Eq. (19) is numerically illustrated in Fig.
1(A), where $\vartheta_{12} \approx 13^\circ$ has typically been
taken. One can clearly see that the value of $\theta$ is sensitive
to that of $(\phi_y - \phi_x)$ and the possibility of $\theta =
45^\circ$ has been ruled out. If $\theta = 30^\circ$
\cite{Giunti2} or $\theta \approx 35.3^\circ$ \cite{Wolfenstein}
is assumed, then the value of $(\phi_y - \phi_x)$ must be
fine-tuned to about $100^\circ$ or $70^\circ$. In these two cases,
the magnitude of ${\cal J}_{\rm MNS}$ can be as large as about
$3\%$, as shown in Fig. 1(B).

We conclude that current experimental data do not favor the
bimaximal mixing pattern of ${\cal F}_\nu$ taken for Eq. (3) or
(8), no matter whether ${\cal F}_\nu$ is real or complex. The
numerical relation $\theta_{12} + \vartheta_{12} \approx 45^\circ$
is achievable, however, if ${\cal F}_\nu$ is allowed to slightly
deviate from the bimaximal mixing form and its free parameters can
properly be fine-tuned. Of course, such a phenomenological
approach is not well motivated from a theoretical point of view.

\vspace{0.3cm}

\framebox{\large\bf 4} ~ What is the most favorite pattern of
${\cal F}_\nu$ in phenomenology? One can straightforwardly answer
this question by calculating ${\cal F}_\nu$ in terms of $V_{\rm
CKM}$ and $V_{\rm MNS}$ through Eq. (3) or (8); namely, ${\cal
F}_\nu = V_{\rm CKM} V_{\rm MNS}$. When the standard
parametrization of $V_{\rm CKM}$ or $V_{\rm MNS}$ in Eq. (12) is
used, one should take account of the arbitrary but nontrivial
phases between $V_{\rm CKM}$ and $V_{\rm MNS}$. In this case
\footnote{Here we have omitted the Majorana phases of CP violation
in $V_{\rm MNS}$, as they can always be rearranged into a pure
phase matrix on the right-hand side of $V_{\rm MNS}$, just like
$Q_\nu$ in Eq. (11). This simplification does not affect our
estimation of the moduli of nine matrix elements of ${\cal
F}_\nu$.},
\begin{equation}
{\cal F}_\nu = V_{\rm CKM}(\vartheta_{12}, \vartheta_{23},
\vartheta_{13}, \delta_{\rm CKM}) \otimes \Omega_\nu \otimes
V_{\rm MNS}(\theta_{12}, \theta_{23}, \theta_{13}, \delta_{\rm
MNS}) \; ,
\end{equation}
where $\Omega_\nu \equiv {\rm Diag}\{e^{i\delta_1}, e^{i\delta_2},
e^{i\delta_3}\}$ is the relative phase matrix between the CKM and
MNS matrices. To be more specific, we fix $(\vartheta_{12},
\vartheta_{23}, \vartheta_{13}) \approx (13^\circ, 2.4^\circ,
0.2^\circ)$ \cite{PDG} and $(\theta_{12}, \theta_{23},
\theta_{13}) \approx (33^\circ, 45^\circ, 3^\circ)$ in our
numerical calculation of ${\cal F}_\nu$. It proves very
instructive and convenient to expand the matrix elements of
$V_{\rm CKM}$ and $V_{\rm MNS}$ in powers of a small parameter
$\lambda \equiv \sin\vartheta_{12}$:
\begin{eqnarray}
V_{\rm CKM}(\vartheta_{12}, \vartheta_{23}, \vartheta_{13},
\delta_{\rm CKM}) & \approx & \left ( \matrix{ \lambda^{0.02} &
\lambda^{1.00} & \lambda^{3.79} e^{-i\delta_{\rm CKM}} \cr
-\lambda^{1.00} & \lambda^{0.02} & \lambda^{2.12} \cr
\lambda^{3.13} - \lambda^{3.79} e^{i\delta_{\rm CKM}} &
-\lambda^{2.12} & \lambda^{0.00} \cr} \right ) \; ,
\nonumber \\
V_{\rm MNS}(\theta_{12}, \;\theta_{23}, \;\theta_{13},
\;\delta_{\rm MNS}) & \approx & \left ( \matrix{ \lambda^{0.12} &
\lambda^{0.41} & ~ \lambda^{1.98} e^{-i\delta_{\rm MNS}} \cr
-\lambda^{0.64} & \lambda^{0.35} & ~ \lambda^{0.23} \cr
\lambda^{0.64} - \lambda^{2.33} e^{i\delta_{\rm MNS}} &
-\lambda^{0.35} & ~ \lambda^{0.23} \cr} \right ) \; ,
\end{eqnarray}
in which only the leading term of each matrix element (except the
(3,1) elements of $V_{\rm MNS}$) is shown. From Eqs. (20) and
(21), we obtain
\small
\begin{equation}
{\cal F}_\nu \approx \Omega_\nu \left ( \matrix{ \lambda^{0.14} (1
- \lambda^{1.50} e^{i\delta_{21}}) & \lambda^{0.43} (1 +
\lambda^{0.92} e^{i\delta_{21}}) & \lambda^{1.23}
(e^{i\delta_{21}} + \lambda^{0.77} e^{-i\delta_{\rm MNS}}) \cr
-\lambda^{0.66} (1 + \lambda^{0.46} e^{-i\delta_{21}}) &
\lambda^{0.37} (1 - \lambda^{1.04} e^{-i\delta_{21}}) &
\lambda^{0.25} (1 + \lambda^{2.10} e^{i\delta_{32}}) \cr
\lambda^{0.64} (1 - \lambda^{1.69} e^{i\delta_{\rm MNS}}) &
-\lambda^{0.35} (1 + \lambda^{2.12} e^{-i\delta_{32}}) &
\lambda^{0.23} (1 - \lambda^{2.12} e^{-i\delta_{32}}) \cr} \right
) \; ,
\end{equation}
\normalsize where $\delta_{ij} \equiv \delta_i - \delta_j$ (for
$i,j = 1,2,3$) has been defined. One can see that the CP-violating
phase $\delta_{\rm CKM}$ does not play a role in this
approximation. The most striking feature of ${\cal F}_\nu$ is that
its (1,3) element does not vanish, unlike the ansatz proposed in
Eq. (11) or (17). This result remains valid even in the
$\theta_{13} = 0$, $\vartheta_{13} =0$ or $\theta_{13} =
\vartheta_{13} =0$ case, simply because $|({\cal F}_\nu)_{13}|$ is
dominated by the product $|(V_{\rm CKM})_{us} (V_{\rm MNS})_{\mu
3}| = \sin\vartheta_{12} \sin\theta_{23} \cos\vartheta_{13}
\cos\theta_{13} \approx \lambda^{1.23}$, as already shown in Eq.
(22). Allowing the unknown phase parameters $\delta_{\rm MNS}$,
$\delta_{21}$ and $\delta_{32}$ to vary between $0$ and $2\pi$, we
arrive at the possible ranges of nine matrix elements of ${\cal
F}_\nu$:
\begin{equation}
|{\cal F}_\nu| \approx \left ( \matrix{ 0.72 \cdot\cdot\cdot 0.90
& ~~~ 0.39 \cdot\cdot\cdot 0.66 ~~~ & 0.11 \cdot\cdot\cdot 0.21
\cr 0.18 \cdot\cdot\cdot 0.56 & 0.45 \cdot\cdot\cdot 0.70 & 0.66
\cdot\cdot\cdot 0.72 \cr 0.35 \cdot\cdot\cdot 0.42 & 0.57
\cdot\cdot\cdot 0.62 & 0.68 \cdot\cdot\cdot 0.74 \cr} \right) \; .
\end{equation}
Note here that the ranges given above are for the {\it individual}
matrix elements. The choice of a specific value for one element
may restrict the magnitudes of some others, because they are
related to one another by the unitarity conditions of ${\cal
F}_\nu$.

Eqs. (22) and (23) illustrate that the correlation between $V_{\rm
MNS}$ and $V_{\rm CKM}$ is rather nontrivial. In particular, the
correlation matrix ${\cal F}_\nu$ is neither real nor bimaximal.
Such a result is of course not a surprise, because Eq. (9) has
clearly indicated that ${\cal F}_\nu$ should not have a too
special pattern. Considering the fact that ${\cal F}_\nu$ consists
of one small rotation angle in its (1,3) sector and two large
rotation angles in its (1,2) and (2,3) sectors, we expect that the
texture of $M_{\rm R}$ must be very nontrivial too.

In the above discussions, we did not take into account possible
quantum corrections to relevant physical parameters between the
scale of grand unified theories $\Lambda_{\rm GUT}$ and the low
energy scales $\Lambda_{\rm LOW}$ at which the mixing angles and
CP-violating phases of $V_{\rm CKM}$ and $V_{\rm MNS}$ can
experimentally be determined. One may roughly classify such
renormalization effects into three categories in the
afore-mentioned seesaw models with grand unification:
\begin{itemize}
\item  The first category is about radiative corrections to the
fermion mass and flavor mixing parameters between the GUT scale
$\Lambda_{\rm GUT}$ and the seesaw scale $M_1$, where $M_1$
denotes the mass of the lightest right-handed neutrino. Typically,
$\Lambda_{\rm GUT} \sim 10^{15\cdot\cdot\cdot 16}$ GeV and $M_1
\sim 10^{8\cdot\cdot\cdot 12}$ GeV hold in most model-building
cases \cite{FX00}. Because three right-handed neutrinos are in
general expected to have a mass hierarchy ($M_1 < M_2 < M_3$), the
threshold effects in the renormalization chain $\Lambda_{\rm GUT}
\rightarrow M_3 \rightarrow M_2 \rightarrow M_1$ are likely to
modify the parameters of neutrino masses and lepton flavor mixing
in a significant way \cite{Mei}. The explicit estimation of such
radiative corrections involves a number of free parameters
\cite{Mei2}, hence its arbitrariness or uncertainty is essentially
out of control. In contrast, quark flavor mixing is expected to be
insensitive to such seesaw threshold effects.

\item  From the seesaw scale $M_1$ to the electroweak scale
$\Lambda_{\rm EW}$ ($\sim 10^2$ GeV), the one-loop renormalization
group equation of the effective neutrino mass matrix $M_\nu$
consists of the contributions from gauge interactions, quark
Yukawa interactions and charged-lepton Yukawa interactions
\cite{Chankowski}. Only the last contribution may affects $V_{\rm
MNS}$, but the quantitative corrections to lepton mixing angles
and CP-violating phases are strongly suppressed in the standard
model and in the supersymmetric standard model with small
$\tan\beta$ (for simplicity, we have assumed that the scale of
supersymmetry breaking is not far away from $\Lambda_{\rm EW}$).
The running behavior of $V_{\rm CKM}$ from $M_1$ to $\Lambda_{\rm
EW}$ is in general dominated by the $t$-quark, $b$-quark and
$\tau$-lepton Yukawa couplings. To be explicit, $\vartheta_{23}$,
$\vartheta_{13}$ and $\delta_{\rm CKM}$ are sensitive to a drastic
change of energy scales, but $\vartheta_{12}$ is not \cite{Babu}.

\item  Within the standard model, the $\Lambda_{\rm EW}$ threshold
effect is negligibly small for both leptons and quarks. Thus it is
unnecessary to take into account the radiative correction to
$V_{\rm MNS}$ and $V_{\rm CKM}$ from $\Lambda_{\rm EW}$ to
$\Lambda_{\rm LOW}$. However, the $\Lambda_{\rm EW}$ threshold
effect may be important in the supersymmetric case
\cite{Chankowski}; e.g., it can be induced by the slepton mass
splitting, which is possible to dominate over the contribution
from the charged-lepton Yukawa couplings. A quantitative
calculation of this threshold effect on $V_{\rm MNS}$ is strongly
model-dependent and arbitrary \cite{Kang}, because it involves
some unknown parameters of supersymmetric particles.
\end{itemize}
It is certainly a challenging task to examine how the relation
$V_{\rm MNS} = V^\dagger_{\rm CKM} {\cal F}_\nu$ gets modified
from the afore-listed radiative corrections, unless a realistic
seesaw model is specifically given and its free parameters are
reasonably assumed. In this sense, we argue that the texture of
${\cal F}_\nu$ is unlikely to take a trivial form (such as the
real bimaximal pattern) and the interesting relations in Eqs. (1)
and (2) are more likely to be a numerical accident.

\vspace{0.3cm}

\framebox{\large\bf 5} ~ In summary, we have investigated the
natural correlation between the MNS lepton mixing matrix $V_{\rm
MNS}$ and the CKM quark mixing matrix $V_{\rm CKM}$ for a class of
seesaw models with grand unification. The correlation matrix
${\cal F}_\nu$ is found to be phenomenologically disfavored to
take the bimaximal mixing form, no matter whether the nontrivial
phases of ${\cal F}_\nu$ are taken into account or not. This
observation turns out to be contrary to some previous arguments
that the CKM matrix might measure the deviation of the MNS matrix
from exact bimaximal mixing. We have also shown that a slight
modification of the bimaximal mixing pattern of ${\cal F}_\nu$ may
allow us to reproduce the quark-lepton complementarity relation,
provided the relevant phase parameters get fine-tuned. Calculating
${\cal F}_\nu$ directly in terms of the mixing angles and
CP-violating phases of $V_{\rm CKM}$ and $V_{\rm MNS}$, we have
demonstrated a striking feature of ${\cal F}_\nu$: its (1,3)
element cannot be vanishing or too small, even if the (1,3)
elements of $V_{\rm MNS}$ and $V_{\rm CKM}$ are taken to be zero.
It is therefore concluded that the texture of ${\cal F}_\nu$ is
rather nontrivial in the seesaw models. We argue that this
conclusion is in general not expected to be changed by possible
quantum corrections and their threshold effects.

\vspace{0.5cm}

The author is grateful to S.K. Kang and J.W. Mei for some helpful
discussions, and to B. Bajc, G. Senjanovic and H. Zhang for
clarifying some misleading points associated with the SO(10)
models. He is also indebted to CSSM at the University of Adelaide,
where this paper was written, for warm hospitality. This work was
supported in part by the National Nature Science Foundation of
China.


\newpage

\begin{figure}
\begin{center}
\vspace{-1cm}
\includegraphics[width=16cm,height=23cm]{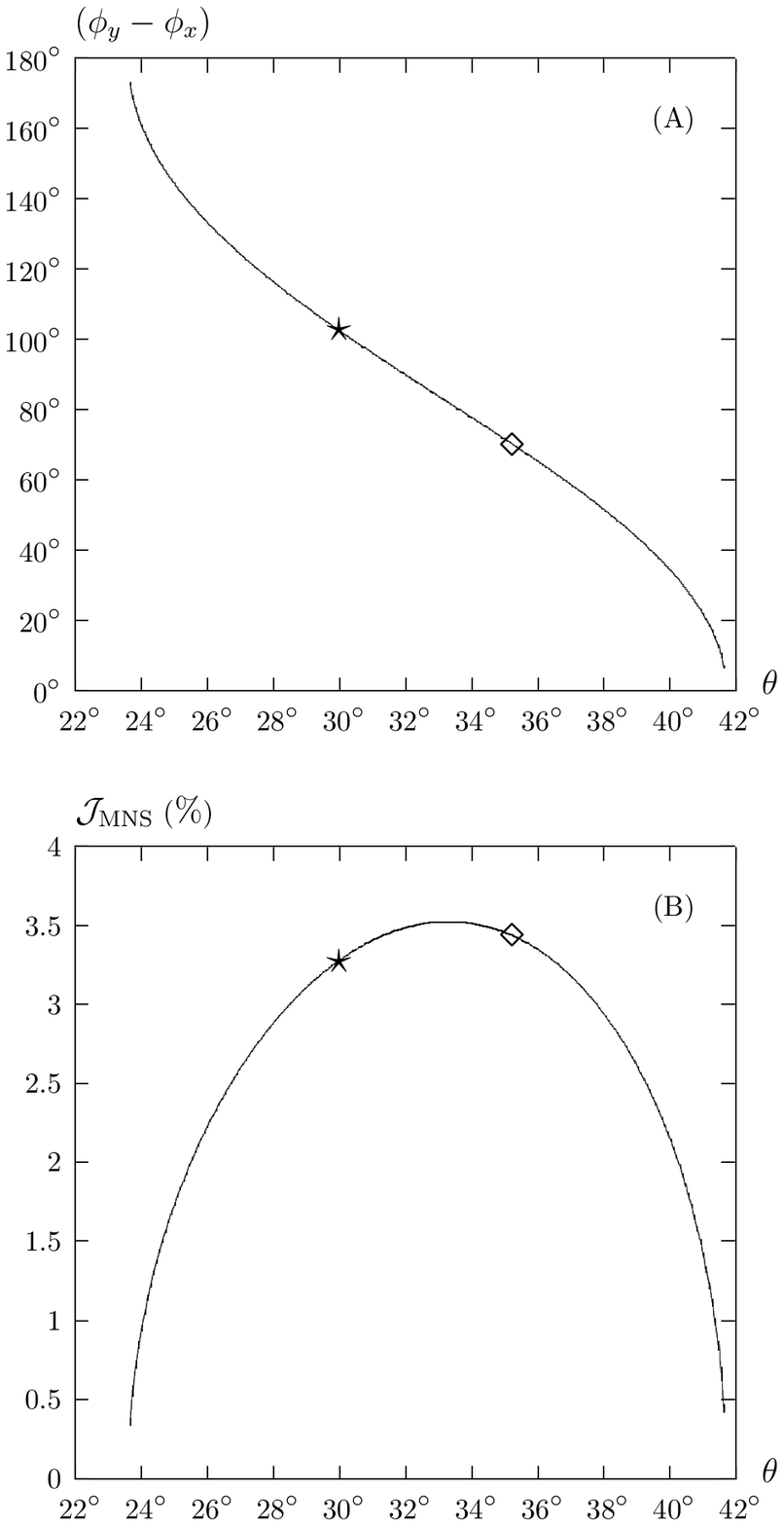}
\vspace{-4.5cm} \caption{(A) numerical illustration of the
correlation between $\theta$ and $(\phi_y - \phi_x)$ for
$\theta_{12} + \vartheta_{12} \approx 45^\circ$ to hold; (B)
numerical dependence of ${\cal J}_{\rm MNS}$ on $\theta$. Here
$\vartheta_{12} \approx 13^\circ$ has been input. The {\it star}
and {\it diamond} points stand for the neutrino mixing ansatz
proposed in Ref. [21] with $\theta =30^\circ$ and that in Ref.
[22] with $\theta \approx 35.3^\circ$, respectively.}
\end{center}
\end{figure}

\end{document}